\documentstyle[12pt,epsfig]{article}
\begin{document}
\centerline{\Large\bf The Event Horizon of The Schwarzschild Black
Hole}
\centerline{\Large\bf in Noncommutative Spaces}
\vspace*{0.050truein}
\centerline{Forough Nasseri\footnote{Email: nasseri@fastmail.fm}}
\centerline{\it Physics Department, Sabzevar University of Tarbiat
Moallem, P.O.Box 397, Sabzevar, Iran}
\centerline{\it Khayyam Planetarium,
P.O.Box 769, Neishabour, Iran}
\begin{center}
(\today)
\end{center}
\begin{abstract}
The event horizon of Schwarzschild black hole is obtained in
noncommutative spaces up to the second order of perturbative calcultions.
Because this type of black hole is non-rotating, to the
first order there is no any effect on the  event horizon due to
the noncommutativity of space. A lower limit for the noncommutativity
parameter is also obtained. As a result, the event horizon in
noncommutative spaces is less than the event
horizon in commutative spaces.\\
\vspace*{2.00mm}
\end{abstract}

Recently, remotivated by string theory \cite{1}, noncommutative spaces
(Moyal plane) have been studied extensively. The noncommutative spaces
can be realized by the coordinate operators satisfying
\begin{equation}
\label{1}
\left[ {\hat x}_i,{\hat x}_j \right] = i \theta_{ij},\;\;\;\;\;{i,j=1,2,3},
\end{equation}
where $\hat x$ are the coordinate operators and $\theta_{ij}$ is the
noncommutativity parameter and is of dimension (length)$^2$.
In noncommutative spaces, the usual product of fields should be replaced
by the star-product
\begin{equation}
\label{2}
f \star g =\exp \left( \frac{i}{2}
\theta_{ij} \frac{\partial}{\partial x^i}
\frac{\partial}{\partial y^j} \right) f(x) g(y)|_{x=y},
\end{equation}
where $f$ and $g$ are two arbitrary infinitely differentiable functions.
In noncommutative spaces we have
\begin{equation}
\label{3}
\left[ {\hat x}_i,{\hat x}_j \right]=i \theta_{ij},\;\;\;
\left[ {\hat x}_i,{\hat p}_j \right]=i \delta_{ij},\;\;\;
\left[ {\hat p}_i,{\hat p}_j \right]=0.
\end{equation}

In \cite{2}, we studied Schwarzschild black holes in
noncommutative spaces up to the first order of perturbative
calculations. Our aim in this letter is to study the effect of the
noncommutativity of space on the event horizon of Schwarzschild black hole
up to the second order of perturbative calculations. We use natural units
in which we set $\hbar=c=k_B=1$ and $G=m_{\rm Pl}^{-2}=l_{\rm Pl}^2$ where
$m_{\rm Pl}$ and $l_{\rm Pl}$ are the Planck mass and the Planck length,
respectively.

The metric of the Schwarzschild black hole is given by
\begin{equation}
\label{4}
ds^2= \left( 1 - \frac{2GM}{r} \right) dt^2 -
\frac{dr^2}{\left( 1 - \frac{2GM}{r} \right)}-
r^2 \left( d \theta^2 +\sin^2 \theta d\phi^2 \right).
\end{equation}
There is an event horizon at $r_h=2GM$.

We introduce the following metric for the Schwarzschild black hole
in the noncommutative spaces \cite{2}
\begin{equation}
\label{5}
ds^2= \left( 1 - \frac{2GM}{\sqrt{{\hat r}{\hat r}}} \right) dt^2 -
\frac{d{\hat r}d{\hat r}}{\left( 1 - \frac{2GM}{\sqrt{{\hat r}{\hat r}}} \right)}-
{\hat r}{\hat r} \left( d \theta^2 +\sin^2 \theta d\phi^2 \right),
\end{equation}
where ${\hat r}$ satisfying (\ref{3}).
The event horizon of the noncommutative metric (\ref{5}) satisfies
the following condition
\begin{equation}
\label{6}
1 - \frac{2GM}{\sqrt{{\hat r}{\hat r}}}=0.
\end{equation}
We note that there is a new coordinate system \cite{3}
\begin{equation}
\label{7}
x_i={\hat x}_i+\frac{1}{2} \theta_{ij} {\hat p}_j,\;\;\;p_i={\hat p}_i,
\end{equation}
where the new variables read the usual canonical commutation relations
\begin{equation}
\label{8}
\left[ x_i,x_j \right] = 0,\;\;\;
\left[ x_i,p_j \right] = i\delta_{ij},\;\;\;
\left[ p_i,p_j \right] = 0.
\end{equation}
So, if we change the variables
${\hat x}_i$ to $x_i$ in (\ref{6}), the singularities of the metric
(\ref{5}) are the solutions of
\begin{equation}
\label{9}
1 - \frac{2GM}{\sqrt{ \left( x_i-\frac{1}{2} \theta_{ij}p_j \right)
\left( x_i-\frac{1}{2}\theta_{ik}p_k \right) }}=0.
\end{equation}
This leads us to
\begin{equation}
\label{10}
1-\frac{2GM}{r} \left( 1 + \frac{x_i \theta_{ij} p_j}{2 r^2}
- \frac{\theta_{ij} \theta_{ik} p_j p_k}{8 r^2} \right)
+ {\cal O} (\theta^3) +...=0,
\end{equation}
where $\theta_{ij}=\frac{1}{2} \epsilon_{ijk} \theta_k$.
Using the identity
\begin{equation}
\label{11}
\epsilon_{ijr} \epsilon_{iks}=
\delta_{jk} \delta_{rs} - \delta_{js} \delta_{rk}
\end{equation}
one can rewrite (\ref{10}) as follows
\begin{equation}
\label{12}
1 - \frac{2 G M}{r} - \frac{GM}{2 r^3} \left[ {\vec L}.{\vec \theta} -
\frac{1}{8} \left( p^2 \theta^2  -({\vec p}.{\vec \theta})^2 \right)
\right] + {\cal O}(\theta^3)+...=0,
\end{equation}
where $L_k=\epsilon_{ijk} x_i p_j$, $p^2={\vec p}.{\vec p}$
and $\theta^2={\vec \theta}.{\vec \theta}$.
If we put $\theta_3=\theta$ and the rest of the $\theta$-components
to zero (which can be done by a rotation or a redefinition of
coordinates), then ${\vec L}.{\vec \theta}=L_z \theta$ and
${\vec p}.{\vec \theta}=p_z \theta$. So, we can rewrite (\ref{12}) as
\begin{equation}
\label{13}
r^3 - 2 G M r^2
- \frac{GM}{2} \left[ L_z \theta - \frac{1}{8}
\left( p^2 - p_z^2 \right) \theta^2 \right]+
{\cal O}(\theta^3)+...=0.
\end{equation}
Using $p^2=p_x^2+p_y^2+p_z^2$ we have $(p^2 - p_z^2) \theta^2
=(p_x^2+p_y^2) \theta^2$. On the other hand, Schwarzschild
black hole is non-rotating, ${\vec L}={\vec 0} \to L_z=0$.
Therefore, we can rewrite (\ref{13}) as follows
\begin{equation}
\label{14}
r^3 - 2 G M r^2
+ \frac{GM}{16} \left( p_x^2 + p_y^2 \right) \theta^2+
{\cal O}(\theta^3)+...=0.
\end{equation}
This equation is a cubic polynomial equation as $x^3+ax^2+bx+c=0$.
In (\ref{14}), we have $b=0$.
Using the three roots $r_1$, $r_2$ and $r_3$ of a cubic polynomial
equation and defining
\begin{eqnarray}
\label{15}
a & \equiv & -2GM,\\
\label{16}
c & \equiv &
\frac{GM}{16} \left( p_x^2 + p_y^2 \right) \theta^2,
\end{eqnarray}
one can obtain the singularities of the metric. The three roots
$r_1$, $r_2$ and $r_3$ of a cubic polynomial equation $x^3+ax^2+bx+c=0$
are given in \cite{4}. The real root of cubic formula (\ref{14}) is the
event horizon of the Schwarzschild black hole in noncommutative
spaces \cite{2}
\begin{eqnarray}
\label{17}
{\hat r}_h& \equiv &-\frac{a}{3} 
+ \left(\frac{-2a^3 - 27c + \sqrt{108a^3c+729c^2}}{54}\right)^{1/3}\nonumber\\
&+& \left(\frac{-2a^3 - 27c - \sqrt{108a^3c+729c^2}}{54}\right)^{1/3}.
\end{eqnarray}
Two other roots of cubic formula (\ref{14}) are not real \cite{4}.
In the case of commutative spaces $c=0$, Eq. (\ref{17}) yields
$r_h=2GM$.

As we know, the event horizon is a real quantity. So the expression under
the square root sign in (\ref{17}) must be positive or zero and
satisfying
\begin{equation}
\label{18}
108 a^3 c + 729 c^2 \geq 0.
\end{equation}
From (\ref{16}), we note that $c \geq 0$. Therefore,
the condition (\ref{18}) yields
\begin{equation}
\label{19}
\frac{512}{27} \left( \frac{G M}{\theta} \right)^2 \leq \left(
p_x^2 + p_y^2 \right).
\end{equation}
Using $p_x^2 + p_y^2 = M^2 (v_x^2 + v_y^2 )$ where $v_x$ and $v_y$ are the
speed of the Schwarzschild black hole along with the $x-$ and $y-$axis
respectively, Eq.(\ref{19}) reads
\begin{equation}
\label{20}
\frac{512}{27} \left( \frac{G}{\theta} \right)^2 \leq \left(
v_x^2 + v_y^2 \right).
\end{equation}
From the principle of special relativity the maximum value of
$v^2=v_x^2+v_y^2$ is unity. So, we have
\begin{equation}
\label{21}
\frac{512}{27} \left( \frac{G}{\theta} \right)^2 \leq 1.
\end{equation}
Inserting $G=l_{\rm Pl}^2$ in (\ref{21}), we are led to
a lower limit for the noncommutativity parameter
\begin{equation}
\label{22}
\theta \geq 4.35\,l_{\rm Pl}^2.
\end{equation}
Let us now compare the event horizon, ${\hat r}_h$, of Schwarzschild black
hole in noncommutative spaces with the event horizon, $r_h=2GM$, of
Schwarzschild black hole in commutative spaces. In doing so,
we rewrite (\ref{17}) as follows
\begin{eqnarray}
\label{23}
{\hat r}_h &=& - \frac{a}{3} + \left[ \left( -\frac{a}{3} \right)^3
-\frac{c}{2} + \frac{c}{2} \left( 1 + \frac{4a^3}{27 c} \right)^{1/2}
\right]^{1/3}\nonumber\\
&+& \left[ \left( -\frac{a}{3} \right)^3 - \frac{c}{2}
- \frac{c}{2} \left( 1 + \frac{4a^3}{27c} \right)^{1/2} \right]^{1/3}.
\end{eqnarray}
From (\ref{18}) we are led to $0 \leq \frac{-4a^3}{27c} \leq 1$
and consequently $0 \leq (1+\frac{4a^3}{27c})^{1/2} \leq 1$.
Therefore, it can be easily seen by (\ref{23}) that ${\hat r}_h$ is
less than $r_h$. In other words, the event horizon of Schwarzschild
black hole in noncommutative spaces is less than the event horizon
of this type of black hole in commutative spaces.

To summarize, the event horizon of Schwarzschild black hole has been
studied in noncommutative spaces up to the second order of perturbative
calculations. Since the noncommutativity parameter is so small in
comparison with the length scales of the system, we have considered the
noncommutative effect as perturbations of the commutative counterpart.

Considering (\ref{13}) one can conclude that additional terms in
perturbative calculations of noncommutativity of space to the first
order and second order are proportional to $L_z$ and $p^2-p_z^2$,
respectively. On the other hand, the Schwarzschild black hole
is non-rotating, ${\vec L}={\vec 0} \to L_z=0$. So, to the first order
of perturbative calculations, there is no any effect on the event horizon
of this type of black hole due to the noncommutativity of space.
To the second order of perturbative calculations, the effect of the
noncommutativity of space appear if the components of the
linear momentum of Schwarzschild black hole not be zero in the $x-y$
plane, see Eq.(\ref{14}).\\
Due to the noncommutativity of space, there is a rotational symmetry
breaking and a preferred direction along with the $z$-axis because
we put $\theta_3=\theta$ and the rest of the $\theta$-components to zero.
As we know, the event horizon is a real quantity. To satisfy this property
of the event horizon, we obtain a lower limit, as given by (\ref{22}), for
the noncommutativity parameter. We also compare the event horizon of
Schwarzschild black hole in noncommutative spaces with the event horizon
of this type of black hole in commutative spaces.
Using (\ref{23}), one can conclude that the event horizon of Schwarzschild
black hole in noncommutative spaces is less than the event horizon
of this type of black hole in commutative spaces.

Our results here can be used for
more studies about the thermodynamics of the Schwarzschild black holes
in noncommutative spaces. This issue will be investigated in a further
communication.\\
{\bf Acknowledgments:}
F.N. thanks Hurieh Husseinian and A.A. Nasseri for noble
helps and also thanks Amir and Shahrokh for truthful helps.


\begin{thebibliography}{99}
\bibitem{1} N. Seiberg and E. Witten, JHEP {\bf 09} (1999) 032;
hep-th/9908142.
\bibitem{2} F. Nasseri, Gen. Relativ. Gravit. {\bf 37} 2223 (2005);
hep-th/0508051.
\bibitem{3} M. Chaichian, M.M. Sheikh-Jabbari and A. Tureanu,
Phys. Rev. Lett. {\bf 86} (2001) 2716; hep-th/0010175.
\bibitem{4} http://planetmath.org/encyclopedia/CubicFormula.html.
\end{thebibliography}
\end{document}